\documentclass{entcs} 
\usepackage{arxiv}
\usepackage{graphicx}
\def\lastname{Grattage}

\makeatletter
\@ifundefined{lhs2tex.lhs2tex.sty.read}%
  {\@namedef{lhs2tex.lhs2tex.sty.read}{}%
   \newcommand\SkipToFmtEnd{}%
   \newcommand\EndFmtInput{}%
   \long\def\SkipToFmtEnd#1\EndFmtInput{}%
  }\SkipToFmtEnd

\newcommand\ReadOnlyOnce[1]{\@ifundefined{#1}{\@namedef{#1}{}}\SkipToFmtEnd}
\usepackage{amstext}
\usepackage{amssymb}
\usepackage{stmaryrd}
\DeclareFontFamily{OT1}{cmtex}{}
\DeclareFontShape{OT1}{cmtex}{m}{n}
  {<5><6><7><8>cmtex8
   <9>cmtex9
   <10><10.95><12><14.4><17.28><20.74><24.88>cmtex10}{}
\DeclareFontShape{OT1}{cmtex}{m}{it}
  {<-> ssub * cmtt/m/it}{}

\DeclareFontShape{OT1}{cmtt}{bx}{n}
  {<5><6><7><8>cmtt8
   <9>cmbtt9
   <10><10.95><12><14.4><17.28><20.74><24.88>cmbtt10}{}
\DeclareFontShape{OT1}{cmtex}{bx}{n}
  {<-> ssub * cmtt/bx/n}{}
\newcommand{\Conid}[1]{\mathit{#1}}
\newcommand{\Varid}[1]{\mathit{#1}}
\newcommand{\anonymous}{\kern0.06em \vbox{\hrule\@width.5em}}

\usepackage{polytable}

\@ifundefined{mathindent}%
  {\newdimen\mathindent\mathindent\leftmargini}%
  {}%

\def\resethooks{%
  \global\let\SaveRestoreHook\empty
  \global\let\ColumnHook\empty}
\newcommand*{\savecolumns}[1][default]%
  {\g@addto@macro\SaveRestoreHook{\savecolumns[#1]}}
\newcommand*{\restorecolumns}[1][default]%
  {\g@addto@macro\SaveRestoreHook{\restorecolumns[#1]}}
\newcommand*{\aligncolumn}[2]%
  {\g@addto@macro\ColumnHook{\column{#1}{#2}}}

\resethooks

\newcommand{\onelinecommentchars}{\quad-{}- }
\newcommand{\commentbeginchars}{\enskip\{-}
\newcommand{\commentendchars}{-\}\enskip}

\newcommand{\visiblecomments}{%
  \let\onelinecomment=\onelinecommentchars
  \let\commentbegin=\commentbeginchars
  \let\commentend=\commentendchars}

\newcommand{\invisiblecomments}{%
  \let\onelinecomment=\empty
  \let\commentbegin=\empty
  \let\commentend=\empty}

\visiblecomments

\newlength{\blanklineskip}
\newcommand{\hsindent}[1]{\quad}% default is fixed indentation

\makeatother
\EndFmtInput

\input{Qcircuit}
\newdimen\proofrulebreadth \proofrulebreadth=.05em
\newdimen\proofdotseparation \proofdotseparation=1.25ex
\newdimen\proofrulebaseline \proofrulebaseline=2ex
\newcount\proofdotnumber \proofdotnumber=3
\let\then\relax
\def\hfi{\hskip0pt plus.0001fil}
\mathchardef\squigto="3A3B
\newif\ifinsideprooftree\insideprooftreefalse
\newif\ifonleftofproofrule\onleftofproofrulefalse
\newif\ifproofdots\proofdotsfalse
\newif\ifdoubleproof\doubleprooffalse
\let\wereinproofbit\relax
\newdimen\shortenproofleft
\newdimen\shortenproofright
\newdimen\proofbelowshift
\newbox\proofabove
\newbox\proofbelow
\newbox\proofrulename
\def\shiftproofbelow{\let\next\relax\afterassignment\setshiftproofbelow\dimen0 }
\def\shiftproofbelowneg{\def\next{\multiply\dimen0 by-1 }%
\afterassignment\setshiftproofbelow\dimen0 }
\def\setshiftproofbelow{\next\proofbelowshift=\dimen0 }
\def\setproofrulebreadth{\proofrulebreadth}

\def\prooftree{% NESTED ZERO (\ifonleftofproofrule)
\ifnum  \lastpenalty=1
\then   \unpenalty
\else   \onleftofproofrulefalse
\fi
\ifonleftofproofrule
\else   \ifinsideprooftree
        \then   \hskip.5em plus1fil
        \fi
\fi
\bgroup% NESTED ONE (\proofbelow, \proofrulename, \proofabove,
\setbox\proofbelow=\hbox{}\setbox\proofrulename=\hbox{}%
\let\justifies\proofover\let\leadsto\proofoverdots\let\Justifies\proofoverdbl
\let\using\proofusing\let\[\prooftree
\ifinsideprooftree\let\]\endprooftree\fi
\proofdotsfalse\doubleprooffalse
\let\thickness\setproofrulebreadth
\let\shiftright\shiftproofbelow \let\shift\shiftproofbelow
\let\shiftleft\shiftproofbelowneg
\let\ifwasinsideprooftree\ifinsideprooftree
\insideprooftreetrue
\setbox\proofabove=\hbox\bgroup$\displaystyle % NESTED TWO
\let\wereinproofbit\prooftree
\shortenproofleft=0pt \shortenproofright=0pt \proofbelowshift=0pt
\onleftofproofruletrue\penalty1
}

\def\eproofbit{% NESTED TWO
\ifx    \wereinproofbit\prooftree
\then   \ifcase \lastpenalty
        \then   \shortenproofright=0pt  % 0: some other object, no indentation
        \or     \unpenalty\hfil         % 1: empty hypotheses, just glue
        \or     \unpenalty\unskip       % 2: just had a tree, remove glue
        \else   \shortenproofright=0pt  % eh?
        \fi
\fi
\global\dimen0=\shortenproofleft
\global\dimen1=\shortenproofright
\global\dimen2=\proofrulebreadth
\global\dimen3=\proofbelowshift
\global\dimen4=\proofdotseparation
\global\count255=\proofdotnumber
$\egroup  % NESTED ONE
\shortenproofleft=\dimen0
\shortenproofright=\dimen1
\proofrulebreadth=\dimen2
\proofbelowshift=\dimen3
\proofdotseparation=\dimen4
\proofdotnumber=\count255
}

\def\proofover{% NESTED TWO
\eproofbit % NESTED ONE
\setbox\proofbelow=\hbox\bgroup % NESTED TWO
\let\wereinproofbit\proofover
$\displaystyle
}%
\def\proofoverdbl{% NESTED TWO
\eproofbit % NESTED ONE
\doubleprooftrue
\setbox\proofbelow=\hbox\bgroup % NESTED TWO
\let\wereinproofbit\proofoverdbl
$\displaystyle
}%
\def\proofoverdots{% NESTED TWO
\eproofbit % NESTED ONE
\proofdotstrue
\setbox\proofbelow=\hbox\bgroup % NESTED TWO
\let\wereinproofbit\proofoverdots
$\displaystyle
}%
\def\proofusing{% NESTED TWO
\eproofbit % NESTED ONE
\setbox\proofrulename=\hbox\bgroup % NESTED TWO
\let\wereinproofbit\proofusing
\kern0.3em$
}

\def\endprooftree{% NESTED TWO
\eproofbit % NESTED ONE
  \dimen5 =0pt% spread of hypotheses
\dimen0=\wd\proofabove \advance\dimen0-\shortenproofleft
\advance\dimen0-\shortenproofright
\dimen1=.5\dimen0 \advance\dimen1-.5\wd\proofbelow
\dimen4=\dimen1
\advance\dimen1\proofbelowshift \advance\dimen4-\proofbelowshift
\ifdim  \dimen1<0pt
\then   \advance\shortenproofleft\dimen1
        \advance\dimen0-\dimen1
        \dimen1=0pt
        \ifdim  \shortenproofleft<0pt
        \then   \setbox\proofabove=\hbox{%
                        \kern-\shortenproofleft\unhbox\proofabove}%
                \shortenproofleft=0pt
        \fi
\fi
\ifdim  \dimen4<0pt
\then   \advance\shortenproofright\dimen4
        \advance\dimen0-\dimen4
        \dimen4=0pt
\fi
\ifdim  \shortenproofright<\wd\proofrulename
\then   \shortenproofright=\wd\proofrulename
\fi
\dimen2=\shortenproofleft \advance\dimen2 by\dimen1
\dimen3=\shortenproofright\advance\dimen3 by\dimen4
\ifproofdots
\then
        \dimen6=\shortenproofleft \advance\dimen6 .5\dimen0
        \setbox1=\vbox to\proofdotseparation{\vss\hbox{$\cdot$}\vss}%
        \setbox0=\hbox{%
                \advance\dimen6-.5\wd1
                \kern\dimen6
                $\vcenter to\proofdotnumber\proofdotseparation
                        {\leaders\box1\vfill}$%
                \unhbox\proofrulename}%
\else   \dimen6=\fontdimen22\the\textfont2 % height of maths axis
        \dimen7=\dimen6
        \advance\dimen6by.5\proofrulebreadth
        \advance\dimen7by-.5\proofrulebreadth
        \setbox0=\hbox{%
                \kern\shortenproofleft
                \ifdoubleproof
                \then   \hbox to\dimen0{%
                        $\mathsurround0pt\mathord=\mkern-6mu%
                        \cleaders\hbox{$\mkern-2mu=\mkern-2mu$}\hfill
                        \mkern-6mu\mathord=$}%
                \else   \vrule height\dimen6 depth-\dimen7 width\dimen0
                \fi
                \unhbox\proofrulename}%
        \ht0=\dimen6 \dp0=-\dimen7
\fi
\let\doll\relax
\ifwasinsideprooftree
\then   \let\VBOX\vbox
\else   \ifmmode\else$\let\doll=$\fi
        \let\VBOX\vcenter
\fi
\VBOX   {\baselineskip\proofrulebaseline \lineskip.2ex
        \expandafter\lineskiplimit\ifproofdots0ex\else-0.6ex\fi
        \hbox   spread\dimen5   {\hfi\unhbox\proofabove\hfi}%
        \hbox{\box0}%
        \hbox   {\kern\dimen2 \box\proofbelow}}\doll%
\global\dimen2=\dimen2
\global\dimen3=\dimen3
\egroup % NESTED ZERO
\ifonleftofproofrule
\then   \shortenproofleft=\dimen2
\fi
\shortenproofright=\dimen3
\onleftofproofrulefalse
\ifinsideprooftree
\then   \hskip.5em plus 1fil \penalty2
\fi
}

\usepackage{color,graphicx} %For comments

\newcommand{\etal}{\emph{et al\ }{}}

\newcommand{\ie}{\emph{i.e.\ }{}}

\newcommand{\FQC}{\mathbf{FQC}}

\newcommand{\FQCo}{\mathbf{FQC}^\circ}

\newcommand{\QML}{\mathbf{QML}}
\newcommand{\QMLo}{\mathbf{QML}^\circ}

\newcommand{\tin}{:}                       %qmlsyn
\newcommand{\vdasho}{\vdash^\circ}         %qmlsyn
\newcommand{\QQ}[1]{{\mathcal Q}_{#1}}
\newcommand{\Qbit}{\QQ{2}}

\newcommand{\Tyid}[1]{\mathbf{#1}}

\newcommand{\sizet}[1]{\abs{#1}}

\newcommand{\rif}{\mathbf{if}}
\newcommand{\rifo}{\mathbf{if}^{\circ}}
\newcommand{\rthen}{\mathbf{then}}
\newcommand{\relse}{\mathbf{else}}

\newcommand{\rsup}{\mathbf{sup}}

\newcommand{\CO}{\mathbf{C}}

\newcommand{\G}{\Gamma}                  %qmlsyn
\newcommand{\D}{\Delta}                  %qmlsyn

\newcommand{\ot}{\otimes}
\newcommand{\abs}[1]{|{#1}|}

\newcommand{\pow}[2]{{#1}^{#2}}

\newcommand{\CNOT}{\textsc{cnot}{}}

\newcommand{\Complex}{\mathbb{C}}

\newcommand{\Cgg}{\multigate{1}{\phi_{\CO}}}
\newcommand{\gCgg}{\ghost{\phi_{\CO}}}
\newcommand{\lab}[1]{\pushT{#1}}

\newcommand{\Super}{\QCat}       %% {\mathbf{Super}}  %{\QQ}        
\newcommand{\Isom}{\QCat^{\circ}}  %%{\mathbf{Isom}}    %{\QQ^\circ}   
\newcommand{\Unit}{\QCat^{\simeq}} %%{\mathbf{Unit}}    %{\QQ^\simeq}   
\newcommand{\QCat}{\mathbf{Q}}

\newcommand{\super}[1]{\widehat{#1}}

\newcommand{\ru}[2]{\vspace{1ex}
\begin{prooftree}
#1 \justifies #2
\end{prooftree}\vspace{1ex}}
\newcommand{\ax}[1]{
\ru{}{#1} }
\newcommand{\Ru}[3]{\vspace{1ex}
\begin{prooftree}
#1 \justifies #2 \using{\rm{#3}}
\end{prooftree}\vspace{1ex}}

\begin{document}
\begin{frontmatter}

\title{An overview of QML with a concrete implementation in Haskell}

  \author{Jonathan Grattage}

  \address{Laboratoire d'Informatique de Grenoble, CNRS, France\\
    \href{mailto:grattage@imag.fr}{\texttt{\normalshape jonathan.grattage@imag.fr}}}

\begin{abstract} 
  This paper gives an introduction to and overview of the functional
quantum programming language QML. The syntax of this language is defined and explained,
along with a new QML definition of the quantum teleport algorithm. The categorical 
operational semantics of QML is also briefly introduced, in the form of annotated 
quantum circuits. This definition leads to a denotational semantics, given in
terms of superoperators. Finally, an implementation in Haskell of the semantics for
QML is presented as a compiler. The compiler takes   
QML programs as input, which are parsed into a Haskell datatype. The output from the compiler is either a
quantum circuit (operational), an isometry (pure denotational) or a superoperator
(impure denotational). Orthogonality judgements and problems with coproducts in QML are also discussed.
\end{abstract}

\thanks{The author gratefully acknowledges Thorsten Altenkirch 
in the development of QML and the operational semantics. Ellie D'Hondt is also
thanked for her constructive suggestions.}

\end{frontmatter}

\section{Introduction and motivation} \label{sec:intro}
Language development for quantum computation is a
rapidly developing research area \cite{GaySJ:quapls},
motivated by the application of established formal reasoning and 
verification techniques within a quantum framework, understanding
the behaviour of quantum computation,
aiding the development of new algorithms and gaining a deeper
understanding of how they work.
This paper discusses the syntax and features of, and gives a compiler for,
a language allowing both classical and quantum control: QML \cite{alti:qml,thesis:jjg}.
The syntax and semantics for QML is a
complete redevelopment of that presented previously \cite{alti:qml}, as
the language has been changed to remove a problematic interpretation of coproducts (section \ref{coprods});
the interpretation of orthogonality has also been updated (section \ref{orth}). 
In addition, in this work the operational semantics is made concrete by a
compiler for QML programs implemented in Haskell.

QML has a syntax similar to other functional languages \cite{GaySJ:quapls},
and is based on strict linear logic: \ie linear logic with contraction, but without
implicit weakening  (in contrast, Selinger's QPL uses affine linear
logic \cite{selinger:qpl}, which allows weakening but not contraction).
QML also integrates reversible and irreversible computations
in a single language, where weakenings (which can give rise to the collapse of
superpositions and entanglement) must be explicit. Contractions are allowed and
are modelled as a form of sharing, analogous to the behaviour of classical functional
languages.  Differences between QML and other languages include the use of a quantum control
operation, provided by the \emph{quantum-if} construct \ensuremath{\mathbf{if}^\circ}. To use the \ensuremath{\mathbf{if}^\circ}
operation, the branches of the computation must be orthogonal (distinguishable), the proof
of which is supplied automatically by the type checker at compilation, and hence the programmer need not
supply the condition, nor need it appear in the syntax of terms. Classical control is provided by
a second \emph{classical-if} construct \ensuremath{\mathbf{if}}, which measures the term being branched over.

QML has both an operational and denotational semantics \cite{thesis:jjg},
supporting formal reasoning principles
with an algebra for equational reasoning and a normal form \cite{eqQML}.
The operational semantics of QML is presented using a categorical
formalisation of the quantum circuit model, which is realised by a compiler
that translates QML programs into `typed' quantum circuits.
The denotational semantics is also implemented, which translates
QML terms, via the operational semantics category \ensuremath{\FQC}, into
either isometries \ensuremath{\Isom} or superoperators \ensuremath{\Super}. 

Recent developments in QML
include a fully operational compiler, with a type inference algorithm for QML terms, and the
automatic derivation and extension of the orthogonality judgements
and circuits required for quantum control, which 
have all been implemented. The implementation of the orthogonality
judgements is such that it can be easily expanded as new rules for 
proving the orthogonality of terms are added to QML.
%%; however,there is no space to detail this here.
This short paper outlines the new syntax of QML and
its semantics.
A new and faithful interpretation of the quantum teleportation algorithm in QML is also
presented as an example of using QML.
Details for using the compiler, and the output it generates, are then provided. Finally, 
the future development of the language, the semantics and compiler,
and the uses of the compiler are described.

\section{The syntax of QML}
This section introduces the syntax for QML (see also ref. \cite{thesis:jjg}).
The symbols \ensuremath{\sigma,\tau} are used to vary over QML types, given by 
\ensuremath{\sigma\mathrel{=}\Tyid{\QQ{1}}\mid \Tyid{\QQ{2}}\mid \sigma\otimes\tau}, where the type constructor is the tensor product \ensuremath{\otimes} corresponding
to a product type and \ensuremath{\Tyid{\QQ{2}}} is a qubit type. $x,y,z$ are used to vary over names. Typing contexts (\ensuremath{\Gamma,\Delta}) are given by
\ensuremath{\Gamma\mathrel{=}\bullet\mid \Gamma,\Varid{x}\mathbin{:}\sigma} where \ensuremath{\bullet} is the empty context. Contexts correspond to functions from
a finite set of variables to types.

Constants $\kappa,\iota\in\Complex$ are also used to define the syntax of expressions. Function
variables are used to refer to previously defined QML programs. The terms of QML consist of those of a
first-order functional language, extended with quantum data, a quantum control structure, and a measurement
operator.  The vector notation \ensuremath{\vec{y}} is used for sequence variables to be measured (weakened). The grammar of QML terms is defined thus: 

\begingroup\par\noindent\advance\leftskip\mathindent\(
\begin{pboxed}\SaveRestoreHook
\column{B}{@{}l@{}}
\column{21}{@{}l@{}}
\column{40}{@{}c@{}}
\column{40E}{@{}l@{}}
\column{44}{@{}l@{}}
\column{45}{@{}l@{}}
\column{49}{@{}l@{}}
\column{59}{@{}l@{}}
\column{61}{@{}l@{}}
\column{67}{@{}l@{}}
\column{77}{@{}l@{}}
\column{78}{@{}c@{}}
\column{78E}{@{}l@{}}
\column{82}{@{}l@{}}
\column{87}{@{}l@{}}
\column{E}{@{}l@{}}
\>[B]{}(\Conid{Variables})\;{}\<[21]
\>[21]{}\Varid{x},\Varid{y},\mathbin{...}{}\<[40]
\>[40]{}\in{}\<[40E]
\>[44]{}\Conid{Vars}{}\<[E]
\\
\>[B]{}(\Conid{Prob}\;\Varid{amplitudes})\;{}\<[21]
\>[21]{}\kappa,\iota,\mathbin{...}{}\<[40]
\>[40]{}\in{}\<[40E]
\>[44]{}{\mathbb C}{}\<[E]
\\
\>[B]{}(\Conid{Patterns})\;{}\<[21]
\>[21]{}\Varid{p},\Varid{q}{}\<[40]
\>[40]{}\mathbin{::=}{}\<[40E]
\>[45]{}\Varid{x}{}\<[49]
\>[49]{}\mid (\Varid{x},\Varid{y}){}\<[E]
\\
\>[B]{}(\Conid{Terms})\;{}\<[21]
\>[21]{}\Varid{t},\Varid{u}{}\<[40]
\>[40]{}\mathbin{::=}{}\<[40E]
\>[45]{}\Varid{x}{}\<[49]
\>[49]{}\mid \pow{\Varid{x}}{\vec{y}}{}\<[61]
\>[61]{}\mid (){}\<[67]
\>[67]{}\mid (\Varid{t},\Varid{u}){}\<[78]
\>[78]{}\mid {}\<[78E]
\>[82]{}\mathbf{let}\;\Varid{p}\mathrel{=}\Varid{t}\;\mathbf{in}\;\Varid{u}{}\<[E]
\\
\>[40]{}\mid {}\<[40E]
\>[44]{}\mathbf{if}\;{}\<[49]
\>[49]{}\Varid{t}\;\mathbf{then}\;\Varid{u}\;\mathbf{else}\;\Varid{u'}{}\<[78]
\>[78]{}\mid {}\<[78E]
\>[82]{}\mathbf{if}^\circ\;{}\<[87]
\>[87]{}\Varid{t}\;\mathbf{then}\;\Varid{u}\;\mathbf{else}\;\Varid{u'}{}\<[E]
\\
\>[40]{}\mid {}\<[40E]
\>[44]{}\pow{\mathbf{qfalse}}{\vec{y}}{}\<[59]
\>[59]{}\mid \pow{\mathbf{qtrue}}{\vec{y}}\mid {}\<[77]
\>[77]{}\kappa\times\Varid{t}\mid \Varid{t}\mathbin{+}\Varid{u}{}\<[E]
\ColumnHook
\end{pboxed}
\)\par\noindent\endgroup\resethooks
Quantum data is modelled using the constructs $\kappa \times t$ and $t + u$.
The term $\kappa \times t$, associates the probability amplitude $\kappa$ 
with the term $t$. The term $t + u$ describes a quantum superposition of $t$ and $u$. Quantum
superpositions are first class values, and can be used in conditional expressions
to provide quantum control. For example: \ensuremath{\mathbf{if}^\circ\;(\mathbf{qtrue}\mathbin{+}\mathbf{qfalse})\;\mathbf{then}\;\Varid{t}\;\mathbf{else}\;\Varid{u}}
evaluates both \ensuremath{\Varid{t}} and \ensuremath{\Varid{u}} and combines
the results in a quantum superposition.
Note that the term
\ensuremath{\lambda_0\times\Varid{t}\mathbin{+}\lambda_1\times\Varid{u}}, where \ensuremath{\Varid{t},\Varid{u}} are not qubits,
is syntactic-sugar for
\ensuremath{\mathbf{if}^\circ\;(\lambda_0\times\mathbf{qtrue}\mathbin{+}\lambda_1\times\mathbf{qfalse})\;\mathbf{then}\;\Varid{t}\;\mathbf{else}\;\Varid{u}}.
The type-checker and orthogonality judgements ensure
that this is a valid operation, by providing a proof that \ensuremath{\Varid{t}} and \ensuremath{\Varid{u}} are orthogonal (distinguishable 
in some way), that their types match, and that they are strict terms (they produce no garbage).

In a quantum control operation, the two branches must be orthogonal, 
otherwise the type system would accept terms that implicitly perform measurements. Without this
restriction ``valid'' programs could be written in QML that are not physically realisable by
a quantum computer. Orthogonality judgements are inferred automatically by static analysis of
terms (see section \ref{orth}).

As an example of superposition formation, the
term \ensuremath{(\frac{1}{\sqrt{2}})\times\mathbf{qfalse}\mathbin{+}(\frac{1}{\sqrt{2}})\times\mathbf{qtrue}} is an equal superposition
of \ensuremath{\mathbf{qfalse}} and \ensuremath{\mathbf{qtrue}}. Normalisation factors that are equal may be omitted.

Finally, a QML program is a sequence of function definitions, where a function definition is given by
\ensuremath{\Varid{f}\;\Gamma\mathrel{=}\Varid{t}\mathbin{:}\tau}. A Haskell-style syntax is used to present program examples. For
example, the QML function below (left) is equivalent to the following Haskell-like code (right):
\begin{eqnarray*}
  \ensuremath{\Varid{f}\;(x_1\mathbin{:}\sigma_1,x_2\mathbin{:}\sigma_2,\mathinner{\ldotp\ldotp},x_n\mathbin{:}\sigma_n)}  &\qquad& \ensuremath{\Varid{f}\mathbin{:}\sigma_1\multimap\sigma_2\mathinner{\ldotp\ldotp}\multimap\sigma_n\multimap\tau}\\
  \ensuremath{\mathrel{=}\Varid{t}\mathbin{:}\tau} \qquad\qquad\qquad\quad    &\qquad& \ensuremath{\Varid{f}\;x_1\;x_2\mathinner{\ldotp\ldotp}x_n\mathrel{=}\Varid{t}}
\end{eqnarray*}

\subsection{QML orthogonality judgements}
\label{orth}
QML has a basic type system that tracks the use of 
variables, preventing them from being weakened inappropriately.
However, the type system still
accepts terms which implicitly perform measurements. As a consequence
QML would accept programs which are not realisable as quantum computations.

Consider the expression 
\ensuremath{\mathbf{if}^\circ\;\Varid{x}\;\mathbf{then}\;\mathbf{qtrue}\;\mathbf{else}\;\mathbf{qtrue}}.
This expression returns \ensuremath{\mathbf{qtrue}} without using any information about \ensuremath{\Varid{x}}.
In order to maintain the invariant that all measurements are explicit,
the type system should reject the above expression. More precisely, 
the expression \ensuremath{\mathbf{if}^\circ\;\Varid{x}\;\mathbf{then}\;\Varid{t}\;\mathbf{else}\;\Varid{u}}
should only be accepted if \ensuremath{\Varid{t}\perp\Varid{u}}.
This notion intuitively ensures that the conditional operator does not
implicitly discard any information about \ensuremath{\Varid{x}} during the evaluation.
The branches of a superposition should also be orthogonal for similar reasons.

Mathematically, two terms, \ensuremath{\Varid{t},\Varid{u}}, are orthogonal if their inner-product
is equal to zero, \ensuremath{\Braket{\Varid{t} \vert \Varid{u}}\mathrel{=}\mathrm{0}}. If this is the case then the judgement
\ensuremath{\Varid{t}\perp\Varid{u}} is true, but if the inner-product yields any other 
value then \ensuremath{\Varid{t}} is not orthogonal to \ensuremath{\Varid{u}}. 
In the presentation of an equational theory for
QML \cite{eqQML}
the orthogonality judgements are replaced by an inner-product
judgement on terms, to much the same effect. However, the inner-product approach
is more informative and flexible, and gives a method of reasoning
about orthogonality, hence in future this method may be adopted
for all terms. 

The following rules give the current QML orthogonality judgements:
{\small \[
 \ax{\ensuremath{\mathbf{qtrue}\perp\mathbf{qfalse}}}
\qquad\vspace{-1em}
 \ax{\ensuremath{\mathbf{qfalse}\perp\mathbf{qtrue}}}
\]\[
 \Ru{\ensuremath{\Varid{t}\perp\Varid{u}}}{\ensuremath{(\Varid{t},\Varid{v})\perp(\Varid{u},\Varid{w})}}{\perp\mathbf{pair}_0}
\qquad\vspace{-.7em}
 \Ru{\ensuremath{\Varid{t}\perp\Varid{u}}}{\ensuremath{(\Varid{v},\Varid{t})\perp(\Varid{w},\Varid{u})}}{\perp\mathbf{pair}_1}
\]\[
\Ru{\ensuremath{\Varid{t}\perp\Varid{u}} \qquad \lambda^*_0\kappa_0 = -\lambda^*_1\kappa_1}
   {\ensuremath{\lambda_0\times\Varid{t}\mathbin{+}\lambda_1\times\Varid{u}} \perp \ensuremath{\kappa_0\times\Varid{t}\mathbin{+}\kappa_1\times\Varid{u}}}
   {\perp\rsup}\vspace{-0.5em}
\]\[
\Ru{\ensuremath{\Varid{t}\perp\Varid{u}} \qquad \lambda^*_0\kappa_0 = -\lambda^*_1\kappa_1}
   {\begin{array}{c}
     \ensuremath{\mathbf{if}^\circ\;(\lambda_0\times\mathbf{qtrue}\mathbin{+}\lambda_1\times\mathbf{qfalse})\;\mathbf{then}\;\Varid{t}\;\mathbf{else}\;\Varid{u}} % \\
     \perp \ensuremath{\mathbf{if}^\circ\;(\kappa_0\times\mathbf{qtrue}\mathbin{+}\kappa_1\times\mathbf{qfalse})\;\mathbf{then}\;\Varid{t}\;\mathbf{else}\;\Varid{u}}
    \end{array}}
   {\perp\rsup\rifo}\vspace{-.6em}
\]\[
\Ru{\ensuremath{\Varid{t}\perp\Varid{u}} \qquad \ensuremath{\Varid{t}\perp\Varid{u'}}}
   {\ensuremath{\Varid{t}\perp\mathbf{if}^\circ\;\Varid{c}\;\mathbf{then}\;\Varid{u}\;\mathbf{else}\;\Varid{u'}}}
   {\perp\ensuremath{\mathbf{if}^\circ}_0}
\qquad
\Ru{\ensuremath{\Varid{t}\perp\Varid{u}} \qquad \ensuremath{\Varid{t}\perp\Varid{u'}}}
   {\ensuremath{\mathbf{if}^\circ\;\Varid{c}\;\mathbf{then}\;\Varid{u}\;\mathbf{else}\;\Varid{u'}\perp\Varid{t}}} 
   {\perp\ensuremath{\mathbf{if}^\circ}_1}
\]}
The first two axioms state that the
basic states of \ensuremath{\mathbf{qtrue}} and \ensuremath{\mathbf{qfalse}} are orthogonal.
The third and fourth rules state that pairs of terms
can be orthogonal, provided that one component of a pair is 
orthogonal to the other pair's corresponding component.
The two $\rsup$ rules state when superpositions
of terms can be orthogonal; the second is a restatement of the 
first, translating superpositions using \ensuremath{\mathbf{if}^\circ}. The final two rules
state that \ensuremath{\mathbf{if}^\circ} statements can be orthogonal if all the component terms are.

The use of \ensuremath{\mathbf{if}^\circ} in QML programs is valid only if
the two branches are orthogonal. Hence, for
the Hadamard operation (section \ref{teleport}), it is required that 
\ensuremath{\mathbin{-}\mathbf{qtrue}\mathbin{+}\mathbf{qfalse}\perp\mathbf{qtrue}\mathbin{+}\mathbf{qfalse}}, with the appropriate
renormalisation. In this case the $\perp\rsup$ rule verifies orthogonality.

The rules for orthogonality given so far are incomplete, and may be extended.
Orthogonality judgements must also be interpreted by the operational semantics, discussed in ref \cite{thesis:jjg}. 

\subsection{Removing coproducts from QML}
\label{coprods}

In a previous version of QML \cite{alti:qml}, here
referred to as \ensuremath{\mathrm{QML}^{\oplus}},
the language included the notion of a tensorial coproduct, 
denoted by \ensuremath{\oplus}. This coproduct has now been removed. 
The types of \ensuremath{\mathrm{QML}^{\oplus}} were generated by \ensuremath{\Tyid{\QQ{1}}}, \ensuremath{\sigma\otimes\tau},
and \ensuremath{\sigma\oplus\tau}. Qubits
were not primitive, but defined as \ensuremath{\Tyid{\QQ{2}}\mathrel{=}\Tyid{\QQ{1}}\oplus\Tyid{\QQ{1}}}.
The coproduct allowed any finite type to be directly represented 
in this way; not just limited to \ensuremath{\Tyid{\QQ{2}}}.
The introduction rules used for \ensuremath{\oplus} were the usual coproduct
rules of a left and a right injection:
{\small
\[
\Ru{\Gamma \vdash^a s : \sigma}
   {\Gamma \vdash^a \mathrm{inl}\ s : \sigma \oplus \tau}
   {+\mathrm{intro}_l}
\vspace{-2em}\qquad
\Ru{\Gamma \vdash^a t : \tau}
   {\Gamma \vdash^a \mathrm{inr}\ t : \sigma \oplus \tau}
   {+\mathrm{intro}_r}
\]}
The coproduct type was interpreted as 
\ensuremath{\sigma\oplus\tau\mathrel{=}\Tyid{\QQ{2}}\otimes\sizet{\sigma\;\sqcup\;\tau}},
where \ensuremath{\sizet{\sigma\;\sqcup\;\tau}} could store a value of either 
\ensuremath{\sizet{\sigma}} or \ensuremath{\sizet{\tau}}, by padding the smaller type.
Using the coproduct and injection rules, \ensuremath{\mathbf{qfalse}} and
\ensuremath{\mathbf{qtrue}} were defined in \ensuremath{\mathrm{QML}^{\oplus}} as $\mathrm{inl} () : \Qbit$
and $\ensuremath{\mathbf{qfalse}} = \mathrm{inr} () : \Qbit$, respectively,
omitting the weakening property of \ensuremath{\mathrm{QML}^{\oplus}}.

Instead of \ensuremath{\mathbf{if}} and \ensuremath{\mathbf{if}^\circ} rules, \ensuremath{\mathrm{QML}^{\oplus}} implemented
two \ensuremath{\oplus}-elimination rules: \ensuremath{\mathbf{case}}, providing  
classical-control (a generalisation of \ensuremath{\mathbf{if}}), and
a quantum-control operation \ensuremath{\mathbf{case}^{\circ}} (generalising \ensuremath{\mathbf{if}^\circ}). 
The quantum (non-measuring) \ensuremath{\oplus}-elimination rule is similar to the standard
coproduct elimination rule, and is given as:
{\small
\[
  \Ru{\begin{array}{lr}
       \G \vdash^a c \tin \sigma\oplus \tau\\
       \D,\ x \tin \sigma \vdasho t \tin \rho\\
       \D,\ y \tin \tau   \vdasho u \tin \rho \quad\quad& t \perp u\\
      \end{array}}
 {\G\ot\D \vdash^a \ensuremath{\mathbf{case}^{\circ}\;\Varid{c}\;\mathbf{of}}\{\ensuremath{\mathbf{inl}\;\Varid{x}} \Rightarrow t\ \vert\ \ensuremath{\mathbf{inr}\;\Varid{y}} \Rightarrow u \} \tin \rho}
 {\oplus elim^\circ}\vspace{-1.75em}
\]}

The non-strict (measuring) case removes the orthogonality
requirement, and does not require sub-terms to be strict.
The \ensuremath{\mathbf{if}^a} rules (\ensuremath{\Varid{a}\in\{\mskip1.5mu \circ,\mathbin{-}\mskip1.5mu\}}; if \ensuremath{\Varid{a}\mathrel{=}\circ} then
the rules are strict, \ie measurement-free)
would then be derived as:
\begingroup\par\noindent\advance\leftskip\mathindent\(
\begin{pboxed}\SaveRestoreHook
\column{B}{@{}l@{}}
\column{3}{@{}l@{}}
\column{9}{@{}l@{}}
\column{12}{@{}l@{}}
\column{18}{@{}l@{}}
\column{21}{@{}l@{}}
\column{27}{@{}l@{}}
\column{30}{@{}c@{}}
\column{30E}{@{}l@{}}
\column{33}{@{}l@{}}
\column{41}{@{}l@{}}
\column{47}{@{}l@{}}
\column{61}{@{}l@{}}
\column{75}{@{}c@{}}
\column{75E}{@{}l@{}}
\column{E}{@{}l@{}}
\>[3]{}\mathbf{if}^a\;{}\<[9]
\>[9]{}\Varid{b}\;{}\<[12]
\>[12]{}\mathbf{then}\;{}\<[18]
\>[18]{}\Varid{t}\;{}\<[21]
\>[21]{}\mathbf{else}\;{}\<[27]
\>[27]{}\Varid{u}{}\<[30]
\>[30]{}\mathrel{=}{}\<[30E]
\>[33]{}\mathbf{case}^a\;{}\<[41]
\>[41]{}\Varid{b}\;\mathbf{of}\;{}\<[47]
\>[47]{}\{\mskip1.5mu \mathbf{inl}\;\anonymous \Rightarrow \Varid{t}{}\<[61]
\>[61]{}\mid \mathbf{inr}\;\anonymous \Rightarrow \Varid{u}{}\<[75]
\>[75]{}\mskip1.5mu\}{}\<[75E]
\ColumnHook
\end{pboxed}
\)\par\noindent\endgroup\resethooks
%%The problem with coproducts, and hence |QMLp|, is that
The branches of a \ensuremath{\mathbf{case}^{\circ}} operation can be of different 
sizes, and this was dealt with in the semantics of \ensuremath{\mathrm{QML}^{\oplus}}
by padding the type of the smaller branch. The padding
of one type in this way can lead to the
garbage becoming entangled with the useful output 
in some way. This happens, for example,
if branching over \ensuremath{\Tyid{\QQ{1}}\otimes\Tyid{\QQ{2}}}. The garbage
created by padding may indirectly
measure the qubit which is being branched over.
Consequently, this approach is not compositional,
and has therefore been rejected.

This version of QML resolves this issue by removing the coproduct.
Qubits are now primitive, as are \ensuremath{\mathbf{if}} and \ensuremath{\mathbf{if}^\circ}. Additionally, the 
strict \ensuremath{\mathbf{if}^\circ} does not allow any garbage to be produced.
Coproducts may be reintroduced, possibly limited to classical types.

\section{An operational semantics for QML}\label{opsem}
The new operational semantics for QML is briefly discussed here, which is
an updated version of that presented in \cite{alti:qml}.
This semantics
is defined by giving a translation from QML terms into morphisms in the category \ensuremath{\FQC}.
FQC morphisms
consist of a reversible quantum circuit \ensuremath{\phi}, the input context \ensuremath{\Gamma}, the output type \ensuremath{\sigma}, and 
the size of the auxiliary heap \ensuremath{\Varid{h}} and any garbage \ensuremath{\Varid{g}}.
Any heap qubits are initially set to $0$ (\ensuremath{\Varid{false}}) in the computational basis, and garbage qubits
can be removed by the partial trace operation at the end of the computation. A full development of
\ensuremath{\FQC} is given in references \cite{thesis:jjg,rev2irrev}.

As an example, the classical \ensuremath{\mathbf{if}} construct is defined
by the following typing rule and operational semantic:
\vspace{-.5em}{\small$$
 \Ru{\begin{array}{l}
     \Gamma \vdash c \tin \Qbit\\
     \Delta \vdash t,u \tin \sigma\\
    \end{array}}
   {\Gamma \ot \Delta \vdash \rif\ c\ \rthen\ t\ \relse\ u \tin \sigma}
   {\rif}
\qquad\qquad\vspace{-2em} 
 \ru{\begin{array}{l}
     \ensuremath{\mathbf{c}\in\FQC\;\Gamma\;\Tyid{\QQ{2}}}\\
     \ensuremath{\mathbf{t}\in\FQC\;\Delta\;\sigma}\\
     \ensuremath{\mathbf{u}\in\FQC\;\Delta\;\sigma}
    \end{array}}
   {\begin{array}{l}
     \ensuremath{\textsc{if}_{\mathrm{Op}}\;\mathbf{c}\;\mathbf{t}\;\mathbf{u}\in\FQC\;(\Gamma\otimes\Delta)\;\sigma}\\
     \ensuremath{\textsc{if}_{\mathrm{Op}}\;\mathbf{c}\;\mathbf{t}\;\mathbf{u}\mathrel{=}( h_{\CO}\mathbin{+} h_{\mathbf{c}}\mathbin{+} h_{\mathbf{t}\vert\mathbf{u}}, g_{\mathbf{c}}\mathbin{+}\mathrm{1},\phi)}
    \end{array}}$$}
\vspace{-.40em}where \ensuremath{\phi} in the operational semantics is the following circuit:{\small
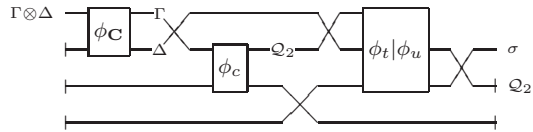
\[
 \Qcircuit@C=1em @R=1em @!R{
  \lstick{\G\ot\D}&\Cgg     &\lab{\G}\qxT&   &\qw                  &\qw           &\qxT&   &\multigate{2}{\phi_t \vert  \phi_u}&    &   &\\
  \lstick{}       &\gCgg\eqw&\lab{\D}\qxB&   &\multigate{1}{\phi_c}&\lab{\Qbit}\qw&\qxB&   &\ghost{\phi_t \vert \phi_u}        &\qxT&   &\rstick{\sigma}\qw\\
                  &\eqw     &\qw         &\qw&\ghost{\phi_c}       &\qxT          &    &\qw&\ghost{\phi_t \vert \phi_u}        &\qxB&   &\rstick{\Qbit}\qwe\\
  \lstick{}       &\eqw     &\qw         &\qw&\qw                  &\qxB          &    &\qw&\qw                                &\qw &\qw&\rstick{}\qwe}
\]}
with $\phi_{\CO}$ as a ``context-splitting'' operation that copies any variables used by both subterms.
$\phi_t \vert  \phi_u$ denotes a conditional circuit which performs $\phi_t$ if the result of the conditional
circuit $c$ is true ($\ket{1})$
and $\phi_u$ if it is false ($\ket{0}$). 

In this way, the semantics of QML is defined recursively over the syntax of terms, and results in a valid FQC morphism for a valid QML term. Given a QML term, the QML compiler follows the operational
semantics and outputs an FQC morphism represented as a \emph{typed circuit}, which can then
either be displayed, exported for use with other programs, or used to directly
implement the program on a quantum-circuit based quantum computer. The FQC morphism can be further
evaluated via the denotational semantics, producing a unitary
matrix \ensuremath{\Unit} (for strict-QML terms without heap), an isometry \ensuremath{\Isom} (for strict-QML terms),
or a superoperator \ensuremath{\Super} (for QML terms that produce garbage). In addition, orthogonality judgements 
and circuits for quantum control and superpositions of terms are automatically 
inferred by the compiler at compile-time, so there can be no orthogonality errors during run-time. The relationships
between QML without measurement (\ensuremath{\QMLo}), full QML (\ensuremath{\QML}), typed quantum circuits (\ensuremath{\FQC}), and isometries
(\ensuremath{\Isom}) and superoperators (\ensuremath{\Super}), is shown in the following diagram: 
{\small
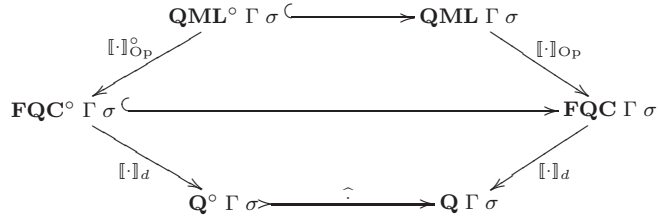
\[
\xymatrix@C=1.5em{
 & \ensuremath{\QMLo\;\Gamma\;\sigma}\ \ar[dl]_{\ensuremath{\llbracket\cdot \rrbracket_{\mathrm{Op}}^\circ}}\ar@{^{(}->}[rrr]& & &     \ensuremath{\QML\;\Gamma\;\sigma}\ar[dr]^{\ensuremath{\llbracket\cdot \rrbracket_{\mathrm{Op}}}}\\
 \ensuremath{\FQCo\;\Gamma\;\sigma}\   \ar[dr]_{\ensuremath{\llbracket\cdot \rrbracket}_d}\ar@{^{(}->}[rrrrr]& & & & & \ensuremath{\FQC\;\Gamma\;\sigma}\ar[dl]^{\ensuremath{\llbracket\cdot \rrbracket}_d}\\
 & \ensuremath{\Isom\;\Gamma\;\sigma}\ \ar@{>->}[rrr]^{\super{.}}           & & &     \ensuremath{\Super\;\Gamma\;\sigma}
}
\]}
where \ensuremath{\llbracket\cdot \rrbracket_{\mathrm{Op}}} denotes the operational semantics, and \ensuremath{\llbracket\cdot \rrbracket_{\mathrm{D}}} denotes the 
additional translation from quantum circuits into the denotational semantics. The full
semantics can be found in reference \cite{thesis:jjg}.

\section{Example: Teleportation in QML}\label{teleport}
The quantum teleportation %is an interesting algorithm that 
describes how to transport a quantum state
using a small amount of classical communication. 
A QML interpretation of the teleportation circuit with
deferred measurement has previously been presented, along with a full description
of the algorithm \cite{thesis:jjg}. However, this circuit relies on the existence
of a quantum channel.
In order to demonstrate and explain the syntax of QML
and the compiler, a new, faithful, implementation of the quantum teleport
algorithm, which explicitly makes use of measurement, is presented.\footnote{The code
is included with the compiler as {\tt teleport.qml}, see section \ref{qmlcomp}.} 
This algorithm is similar to that developed
by Selinger and Valiron \cite{SelingerP:lamcqc}, and also in reference
\cite{Alejandro:meas}, which includes a relevant
discussion of teleportation, both with and without deferred measurement. These
examples show the elegance of allowing quantum control via the quantum
\ensuremath{\mathbf{if}^\circ} construct (\ensuremath{\Conid{CNot}}) and term superpositions (\ensuremath{\Conid{Epr}}), in writing functions. For simple one qubit
functions (\ensuremath{\Conid{Had}}), the branches of the quantum control are the columns from the unitary matrices that
describe the operation. A simple measurement operation, using classical control, is also defined in the following example
which implements the teleportation algorithm (\ensuremath{\Conid{Tele}}):

{\small
\begingroup\par\noindent\advance\leftskip\mathindent\(
\begin{pboxed}\SaveRestoreHook
\column{B}{@{}l@{}}
\column{6}{@{}l@{}}
\column{8}{@{}l@{}}
\column{9}{@{}l@{}}
\column{11}{@{}l@{}}
\column{12}{@{}l@{}}
\column{15}{@{}l@{}}
\column{17}{@{}l@{}}
\column{18}{@{}l@{}}
\column{20}{@{}l@{}}
\column{24}{@{}l@{}}
\column{25}{@{}l@{}}
\column{26}{@{}l@{}}
\column{33}{@{}l@{}}
\column{34}{@{}l@{}}
\column{36}{@{}l@{}}
\column{37}{@{}l@{}}
\column{42}{@{}l@{}}
\column{54}{@{}l@{}}
\column{61}{@{}l@{}}
\column{E}{@{}l@{}}
\>[B]{}\Conid{Had},\Conid{Qnot},\Conid{Meas}{}\<[18]
\>[18]{}\in\Conid{Q2}\multimap\Conid{Q2}{}\<[E]
\\
\>[B]{}\Conid{Had}\;\Varid{b}{}\<[9]
\>[9]{}\mathrel{=}\mathbf{if}^\circ\;\Varid{b}\;{}\<[18]
\>[18]{}\mathbf{then}\;{}\<[24]
\>[24]{}\mathbf{qfalse}\mathbin{+}{}\<[34]
\>[34]{}\mathbin{-}\mathbf{qtrue}{}\<[42]
\>[42]{}\mbox{\onelinecomment  The Hadamard operation}{}\<[E]
\\
\>[18]{}\mathbf{else}\;{}\<[24]
\>[24]{}\mathbf{qfalse}\mathbin{+}{}\<[34]
\>[34]{}\mathbf{qtrue}{}\<[E]
\\[\blanklineskip]
\>[B]{}\Conid{Qnot}\;\Varid{b}{}\<[9]
\>[9]{}\mathrel{=}\mathbf{if}^\circ\;\Varid{b}\;\mathbf{then}\;\mathbf{qfalse}\;\mathbf{else}\;\mathbf{qtrue}{}\<[42]
\>[42]{}\mbox{\onelinecomment  The Not operation (Pauli X)}{}\<[E]
\\[\blanklineskip]
\>[B]{}\Conid{Meas}\;\Varid{b}{}\<[9]
\>[9]{}\mathrel{=}\mathbf{if}\;\Varid{b}\;\mathbf{then}\;\mathbf{qtrue}\;\mathbf{else}\;\mathbf{qfalse}{}\<[42]
\>[42]{}\mbox{\onelinecomment  A measurement operator, using \ensuremath{\mathbf{if}}}{}\<[E]
\\[\blanklineskip]
\>[B]{}\Conid{CNot}{}\<[11]
\>[11]{}\in\Conid{Q2}\multimap\Conid{Q2}\multimap\Conid{Q2}\otimes\Conid{Q2}{}\<[42]
\>[42]{}\mbox{\onelinecomment  A quantum $\CNOT$ operation}{}\<[E]
\\
\>[B]{}\Conid{CNot}\;\Varid{s}\;\Varid{t}{}\<[11]
\>[11]{}\mathrel{=}\mathbf{if}^\circ\;\Varid{s}\;{}\<[20]
\>[20]{}\mathbf{then}\;{}\<[26]
\>[26]{}(\mathbf{qtrue},\Conid{Qnot}\;\Varid{t}){}\<[E]
\\
\>[20]{}\mathbf{else}\;{}\<[26]
\>[26]{}(\mathbf{qfalse},\Varid{t}){}\<[E]
\\[\blanklineskip]
\>[B]{}\Conid{Epr}{}\<[6]
\>[6]{}\in\Conid{Q2}\otimes\Conid{Q2}{}\<[42]
\>[42]{}\mbox{\onelinecomment  The EPR pair, $\ket{00}+\ket{11}$}{}\<[E]
\\
\>[B]{}\Conid{Epr}{}\<[6]
\>[6]{}\mathrel{=}(\mathbf{qtrue},\mathbf{qtrue})\mathbin{+}(\mathbf{qfalse},\mathbf{qfalse}){}\<[E]
\\[\blanklineskip]
\>[B]{}\Conid{Bmeas}{}\<[12]
\>[12]{}\in\Conid{Q2}\multimap\Conid{Q2}\multimap\Conid{Q2}\otimes\Conid{Q2}{}\<[42]
\>[42]{}\mbox{\onelinecomment  The Bell-measurement operation}{}\<[E]
\\
\>[B]{}\Conid{Bmeas}\;\Varid{x}\;\Varid{y}{}\<[12]
\>[12]{}\mathrel{=}{}\<[15]
\>[15]{}\mathbf{let}\;(\Varid{x'},\Varid{y'})\mathrel{=}\Conid{CNot}\;\Varid{x}\;\Varid{y}{}\<[E]
\\
\>[15]{}\mathbf{in}\;(\Conid{Meas}\;(\Conid{Had}\;\Varid{x'}),\Conid{Meas}\;\Varid{y'}){}\<[E]
\\[\blanklineskip]
\>[B]{}\Conid{U}{}\<[9]
\>[9]{}\in\Conid{Q2}\multimap\Conid{Q2}\otimes\Conid{Q2}\multimap\Conid{Q2}{}\<[42]
\>[42]{}\mbox{\onelinecomment  The unitary correction operations}{}\<[E]
\\
\>[B]{}\Conid{U}\;\Varid{q}\;\Varid{xy}{}\<[9]
\>[9]{}\mathrel{=}{}\<[12]
\>[12]{}\mathbf{let}\;(\Varid{x},\Varid{y})\mathrel{=}\Varid{xy}\;\mathbf{in}\;\mathbf{if}\;\Varid{x}\;{}\<[36]
\>[36]{}\mathbf{then}\;{}\<[42]
\>[42]{}(\mathbf{if}\;\Varid{y}\;\mathbf{then}\;{}\<[54]
\>[54]{}U_{11}\;\Varid{q}\;{}\<[61]
\>[61]{}\mathbf{else}\;U_{10}\;\Varid{q}){}\<[E]
\\
\>[36]{}\mathbf{else}\;{}\<[42]
\>[42]{}(\mathbf{if}\;\Varid{y}\;\mathbf{then}\;{}\<[54]
\>[54]{}U_{01}\;\Varid{q}\;{}\<[61]
\>[61]{}\mathbf{else}\;\Varid{q}){}\<[E]
\\
\>[B]{}U_{01},U_{10},U_{11}\in\Conid{Q2}\multimap\Conid{Q2}{}\<[E]
\\
\>[B]{}U_{01}\;\Varid{x}{}\<[8]
\>[8]{}\mathrel{=}\mathbf{if}^\circ\;\Varid{x}\;\mathbf{then}\;{}\<[25]
\>[25]{}\mathbf{qfalse}\;{}\<[33]
\>[33]{}\mathbf{else}\;\mathbf{qtrue}{}\<[E]
\\
\>[B]{}U_{10}\;\Varid{x}{}\<[8]
\>[8]{}\mathrel{=}\mathbf{if}^\circ\;\Varid{x}\;\mathbf{then}\mathbin{-}{}\<[25]
\>[25]{}\mathbf{qtrue}\;{}\<[33]
\>[33]{}\mathbf{else}\;\mathbf{qfalse}{}\<[E]
\\
\>[B]{}U_{11}\;\Varid{x}{}\<[8]
\>[8]{}\mathrel{=}\mathbf{if}^\circ\;\Varid{x}\;\mathbf{then}\mathbin{-}{}\<[25]
\>[25]{}\mathbf{qfalse}\;{}\<[33]
\>[33]{}\mathbf{else}\;\mathbf{qtrue}{}\<[E]
\\[\blanklineskip]
\>[B]{}\mbox{\onelinecomment  The quantum teleportation algorithm}{}\<[E]
\\
\>[B]{}\Conid{Tele}{}\<[9]
\>[9]{}\in\Conid{Q2}\multimap\Conid{Q2}{}\<[E]
\\
\>[B]{}\Conid{Tele}\;\Varid{q}{}\<[9]
\>[9]{}\mathrel{=}{}\<[12]
\>[12]{}\mathbf{let}\;{}\<[17]
\>[17]{}(\Varid{a},\Varid{b}){}\<[24]
\>[24]{}\mathrel{=}\Conid{Epr}{}\<[37]
\>[37]{}\mbox{\onelinecomment  \ensuremath{\Varid{a}} is given to Alice, \ensuremath{\Varid{b}} to Bob}{}\<[E]
\\
\>[17]{}\Varid{f}{}\<[24]
\>[24]{}\mathrel{=}\Conid{Bmeas}\;\Varid{q}\;\Varid{a}{}\<[37]
\>[37]{}\mbox{\onelinecomment  Result of Alice's Bell-measurement is classical data}{}\<[E]
\\
\>[12]{}\mathbf{in}\;\Conid{U}\;\Varid{b}\;\Varid{f}{}\<[37]
\>[37]{}\mbox{\onelinecomment  Bob applies \ensuremath{\Conid{U}} to his qubit, using classical data \ensuremath{\Varid{f}} }{}\<[E]
\ColumnHook
\end{pboxed}
\)\par\noindent\endgroup\resethooks
}\vspace{-1em}
\section{The  QML compiler}\label{qmlcomp}
This section briefly describes the design and operation of the Haskell
QML compiler, which can be found on the project website.\footnote{Instructions for
downloading the QML compiler can be found on the QML compiler 
website \url{http://fop.cs.nott.ac.uk/qml/compiler}. The Haskell 
compiler GHC is also required. To use the compiler, the QML
system must be loaded into GHC via the command
``\ensuremath{\Varid{ghci}\;\Varid{qml}}''. This will initialise the system with all the
modules required to interpret and evaluate QML programs.}
The objective is to take  a file
containing a QML program, consisting of QML function definitions, and
output an annotated (typed) quantum circuit which realises the QML program as
an \ensuremath{\FQC} object.
This is achieved by implementing the operational semantics in Haskell. Additionally,
the circuit produced by the compiler can be further processed to produce either
a unitary matrix (\ensuremath{\Unit}), isometry (\ensuremath{\Isom}), or superoperator (\ensuremath{\Super}),
as appropriate, giving an implementation of the denotational semantics of QML,
as factored through the operational semantics.

The compiler has a modular design, giving a clear logical structure. For example,
the \ensuremath{\Conid{QOrth}} module contains all the code for generating the orthogonality judgements and circuits, while 
\ensuremath{\Conid{QCirc}} contains the definition of the circuit datatypes and associated functions. The compiler
exploits advanced Haskell features, such as monads and pattern matching, and making use of the ideas
put forward by Vizzotto \etal \cite{alti:qeff}. The operational semantics is realised in the 
\ensuremath{\Conid{QComp}} (\emph{Q}ML \emph{Comp}ilation) module. 

\sloppy
To compile QML functions into the operational
semantics (\ensuremath{\FQC} morphisms, represented as typed circuits), the ``run typed
circuit'' command is used: \ensuremath{\Varid{runTC}\;\text{\tt \char34 \#filename\char34}\;\text{\tt \char34 \#function\char34}}. For example,
to evaluate the typed circuit of the function \ensuremath{\Conid{Epr}} (from section \ref{teleport})
the command is \ensuremath{\Varid{runTC}\;\text{\tt \char34 teleport.qml\char34}\;\text{\tt \char34 Epr\char34}}. This outputs the following 
typed circuit (as a Haskell datatype):
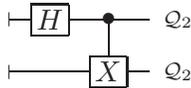
\[
\Qcircuit@C=.75em @R=.75em {
 &\gate{H}\eqw&\ctrl{1}&\rstick{\Qbit}\qw\\
 &\eqw        &\gate{X}&\rstick{\Qbit}\qw
}
\]
where there is no input context, the heap is two qubits (marked $\vdash$), and 
a pair of qubits are the output. \ensuremath{\Conid{H}} and \ensuremath{\Conid{X}} are the Hadamard and Pauli-\ensuremath{\Conid{X}} operations.
As heap is initialised to \ensuremath{\mathbf{qfalse}}, this
circuit describes the function $\ket{00}\ensuremath{\to }\ket{00}+\ket{11}$, producing an EPR pair.
The compilation of a QML term into a circuit is efficient; each QML term
is recursively translated directly into an FQC morphism (an annotated circuit),
as described in section \ref{opsem}. 
No quantum computation is simulated, so this does not 
effect the efficiency of the compiler - it is a simple recursive translation into circuits.
The output is further
optimised after compilation, by collapsing identities and permutations and
other simple circuit manipulations which do not effect the action of the circuit (see \ensuremath{\Conid{QCirc}}).

There are three main options for further evaluation of a QML term: 
\begin{itemize}
  \item The QML term could be evaluated to a unitary matrix (\ensuremath{\Unit}), interpreting
        only the reversible circuit \ensuremath{\phi} from the \ensuremath{\FQC} morphism. As this option
        classically computes the full mathematical interpretation of the program it is
        inherently inefficient. The output is
        actually a triple \ensuremath{(\Varid{h},\Varid{g},\phi)\in(\Conid{Int},\Conid{Int},}Unit\ensuremath{)}, where \ensuremath{\Varid{h},\Varid{g}} gives the size of
        any heap required or garbage produced.
        The command for producing a unitary matrix is \ensuremath{\Varid{runM}};
  \item The function could be evaluated to an isometry (\ensuremath{\Isom}),
        which initialises any required heap, and is the full description
        for terms that produce no garbage. This option is 
        no less efficient than the \ensuremath{\Varid{runM}} option, and is 
        the preferred evaluation option. The output is
        actually a pair \ensuremath{(\Varid{g},\phi)\in(\Conid{Int},\Isom)}, where \ensuremath{\Varid{g}} gives the size of
        any garbage (if the QML function is impure). 
        The command for evaluating to an isometry is \ensuremath{\Varid{runI}};
  \item Thirdly, the QML term can be interpreted as a superoperator (\ensuremath{\Super}),
        which initialises heap and traces out any garbage, using the command
        \ensuremath{\Varid{runS}}. This option is substantially less efficient than the previous two options,
        as the state space is doubled and the partial trace is a computationally 
        expensive operation.
\end{itemize}

Together, the options \ensuremath{\Varid{runI}} and \ensuremath{\Varid{runS}} give an interpretation of the denotational semantics
of QML factored through the category \ensuremath{\FQC}, as shown in the diagram in
section \ref{opsem}. A direct implementation of the denotational
semantics, without using the operational semantics, is an extension
currently being developed. Please refer to the project website for full details.

\section{Conclusions and further work}
\label{sec:concl}
This paper introduces the language QML and presents its semantics with a compiler
and example programs. The semantics and compiler 
give a realisable interpretation of QML programs as quantum circuits in a formal,
categorical, setting. 
The semantics can be extended in many ways, such as expanding the current
orthogonality judgements, or by the addition of non-linear, classical, data. The algebra
for the pure fragment of QML \cite{eqQML} is currently being extended to include measurement,
following work on van Tonder's quantum lambda calculus \cite{Alejandro:meas}. 
It would be instructive to implement this algebra, especially the normal form, as 
part of the QML compiler. Future possibilities for the development of the language
also including developing a notion of higher-order functions for QML, and adding iteration to
the language.

The development of QML and the compiler is an ongoing
project which has already reached a functional state.
As the language and semantics evolve, extensions and new features
can be incorporated into the compiler; which also
provides a useful testbed for the development of new language features and capabilities.
For example, an extension of the orthogonality circuits given in \cite{thesis:jjg}
was developed using the compiler in this way. 
The compiler also facilitates the testing and development of new QML algorithms, such 
as the described teleportation algorithm. It has also been useful in allowing others
to experiment with quantum programming and get immediate feedback on the behaviour of
their functions, in a style that is familiar to computer scientists, logicians, and physicists  
 with functional programming experience.

Further extensions to the compiler include adding the ability to export typed circuits as images,
or in notation compatible with tools such as MatLab and Mathematica. Possible relationships with the measurement
calculus, the Haskell QIO monad \cite{txa:QIO}, and other formalisms are being studied, and may provide new insights.
This will lead to new features being developed, such as basis independence, and further useful abstractions.

\bibliographystyle{entcs}
\bibliography{local}

\end{document}